\RequirePackage{fix-cm}
\documentclass[twocolumn,epjc3]{svjour3}
\smartqed  
\RequirePackage{graphicx}

\usepackage{amsmath}
\usepackage{xcolor}

\journalname{Eur. Phys. J. C}
\begin{document}

\title{Investigation on radiation generated by Sub-GeV electrons in ultrashort Si and Ge bent crystals}

\author{L. Bandiera\thanksref{addr1} \and A. Sytov\thanksref{addr1} \and D. De Salvador \thanksref{addr2} \and A. Mazzolari \thanksref{addr1} \and E. Bagli \thanksref{addr1} \and R. Camattari \thanksref{addr1,addr3} \and  S. Carturan \thanksref{addr2} \and C. Durighello \thanksref{addr1,addr2} \and G. Germogli \thanksref{addr1} \and V. Guidi \thanksref{addr1,addr3} \and P. Klag \thanksref{addr4} \and W. Lauth \thanksref{addr4} \and G. Maggioni \thanksref{addr2} \and V. Mascagna \thanksref{addr7,addr8} \and  M. Prest \thanksref{addr7,addr8} \and M. Romagnoni \thanksref{addr1,addr5} \and M. Soldani \thanksref{addr1,addr3} \and V.V. Tikhomirov \thanksref{addr6} \and E. Vallazza \thanksref{addr7}}




\institute{INFN Section of Ferrara, Via Saragat 1, 44122 Ferrara, Italy \label{addr1}
           \and
           INFN Laboratori Nazionali di Legnaro, Viale dell'Universit\`{a} 2, 35020 Legnaro, Italy\\
           Department of Physics, University of Padova, Via Marzolo 8, 35131 Padova, Italy \label{addr2}
           \and
           Department of Physics and Earth Sciences, University of Ferrara, Via Saragat 1, 44122 Ferrara, Italy \label{addr3}
           \and
           Institut f\"{u}r Kernphysik der Universit\"{a}t Mainz, D-55099 Mainz, Germany \label{addr4}
           \and
           INFN Section of Milano Bicocca, Piazza della Scienza, 3 I20126 Milano, Italy \label{addr7}+ù
           \and
           University of Insubria, Dipartimento di Scienza e Alta Tecnologia (DiSAT), Via Valleggio 11, 22100 Como, Italy \label{addr8}
           \and
           Department of Physics, University of Milan, Via Giovanni Celoria 16, 20133 Milano, Italy \label{addr5}
           \and
           Institute for Nuclear Problems, Belarusian State University, Bobruiskaya 11, 220030 Minsk, Belarus \label{addr6}
}

\date{Received: date / Accepted: date}

\maketitle

\begin{abstract}
{We report on the measurements of the spectra of gamma radiation generated by 855 MeV electrons in bent silicon and germanium crystals at MAMI (MAinzer MIkrotron). The crystals were 15 $\mu m$ thick along the beam direction to ensure high deflection efficiency. Their (111) crystalline planes were bent by means of a piezo-actuated mechanical holder, which allowed to remotely change the crystal curvature. In such a way it was possible to investigate the radiation emitted under planar channeling and volume reflection as a function of the curvature of the crystalline planes.
We show that using volume reflection, one can produce intense gamma radiation with comparable intensity but higher angular acceptance than for channeling. We studied the trade-off between radiation intensity and angular acceptance at different values of the crystal curvature. The measurements of radiation spectra have been carried out for the first time in bent Germanium crystals. In particular, the intensity of radiation in the Ge crystal  is higher than in the Si one due to the higher atomic number, which is important for the development of the X-ray and gamma radiation sources based on higher-Z deformed crystals, such as crystalline undulator.}

\keywords{Channeling \and Volume reflection \and Bent crystal \and Channeling radiation \and Radiation generation}
\PACS{52.59.-f \and 61.80.Fe \and 78.70.-g \and 61.85.+p \and 29.27.Eg \and 29.20.-c}
\end{abstract}

\section{Introduction}

Since the '60s, it has been known that electrons and positrons interacting with strong crystalline fields can generate more intense radiation with respect to standard bremsstrahlung \cite{TerMikaelian,BaierKatkov,Akhiezer:Shulga}. Indeed, the electric field of a crystalline plane or axis can be very high, reaching the values of $10^9$-$10^{12}$ V/cm. For comparison, such a value cannot be artificially created by magnets or by electrostatic deflectors. The first effect discovered was the \textit{coherent bremsstrahlung} (CB) \cite{TerMikaelian}, which occurs when the momentum transferred by the electron/positron to the medium matches a reciprocal lattice vector. It manifests itself for small incidence angle of the charged particles w.r.t. the crystal planes or axes. On the other hand, if the incidence angle of electron/positron is lower than the critical value, namely the Lindhard angle $\theta_L= \sqrt{2U_0/E}$ \cite{Lin}, being $U_0$ the planar/axial potential well depth and $E$ the initial particle energy, the particles are forced to move almost parallel to the planes, oscillating between/around them, depending on their charge sign. This is the channeling effect \cite{Lin}, while the radiation produced during such a motion is called channeling radiation (CR) \cite{Kumakhov197617,Baryshevsky1980}. CR is typically softer but more intense than the CB. However, CR is limited by a low angular acceptance, i.e. within the Lindhard angle, which decreases with the square root of the particle energy. Another limit of CR is the process of dechanneling \cite{Beloshitsky,Biryukov}, i.e. the process for which charged particles escape from the channeling condition due to scattering of the particles on nuclei and electrons of the crystal. Since the intensity of CR depends on the number of charged particles fulfilling the channeling condition, the crystal length must be at most comparable with the dechanneling length, namely the typical length after which a channeled particle suffers the dechanneling effect.

In the last decade, channeling and the related effects have been investigated in bent crystals, mostly for beam steering for applications in collimation and extraction in hadronic accelerators \cite{PhysRevSTAB.5.043501,Scandale2016129,Lie,WIENANDS201711,AfoninJETP,Scandale2013182}. Recently, such investigations were extended to the cases of electrons and positrons \cite{PhysRevA.79.012903,PhysRevLett.114.074801,Wistisen2017,Ban2,Ban,Ban3,SYTOV2017,PhysRevLett.112.135503}, for which the processes of radiation emission are modified by the presence of the crystal curvature. In fact, a bent crystal possesses larger variety of radiation effects than a straight one. An effect that occurs only in bent crystals is the so-called volume reflection (VR) \cite{Taratin1987425}, where the particles are reflected by a bent crystal plane if their incidence angle is a little bit higher than the Lindhard angle. Since VR occurs for over-barrier particles, i.e. for not channeled particles, it is not affected by dechanneling, thereby VR has an higher deflection efficiency than channeling, while its deflection angle is comparable to $\theta_L$. As a consequence of VR there is emission of radiation \cite{Chesnokov_2008,Gui}, which has a higher angular acceptance w.r.t. CR, being equal to the bending angle of the crystal \cite{Ban2,Ban,Ban3}, while the radiation intensity is comparable for both the effects. In particular, the radiation accompanying VR could be exploited to realize intense radiation sources with poor beam emittance \cite{Ban3}.

Another mechanism of radiation generation in a bent crystal is by means of a crystalline undulator (CU) \cite{Baryshevsky1980,korol2013channeling,undu,procChan2018}.This device implies channeling in a periodically bent crystal. The use of a CU would allow to increase the radiation intensity and to decrease the radiation cone angle w.r.t. CR. However, the application of a CU is limited by channeling acceptance and dechanneling. Actually, the investigation of the radiation generation in a bent crystal is a necessary step for the development of a CU. In particular, one should study how the spectral intensity depends on the crystal alignment w.r.t. the beam, as well as on the crystal curvature. Additionally, one should develop and validate a simulation model, predicting the radiation spectra.

An important factor for the realization of an innovative crystal-based radiation source is the choice of the crystalline material. The most frequently used material for channeling in bent crystals is silicon, since it is well established to produce high quality bent Si monocrystals, the lattice of which provides high electric fields. However, even higher electric fields can be produced by materials of higher Z, for instance germanium, which can be manufactured by similar procedures and with the similar quality as silicon crystals \cite{SYTOV2017,DeSalvador2011,DeSalvador2013}. Therefore, higher-Z crystals are very challenging to increase radiation intensity \cite{SHEN201826}.

In this paper, we show the experimental results obtained via CR and VR radiation in ultrashort Si and Ge bent crystals at the Mainz Mikrotron (MAMI), using an 855 MeV electron beam. We compared the experimental spectra with simulations as a function of curvature radius and crystal alignment, both for Si and Ge samples. In particular, the radiation in a Ge bent crystal and its dependence on the crystal curvature are studied for the first time.

\section{Experimental setup}

\begin{figure*}
\begin{center}
    \includegraphics[width=0.8\textwidth]{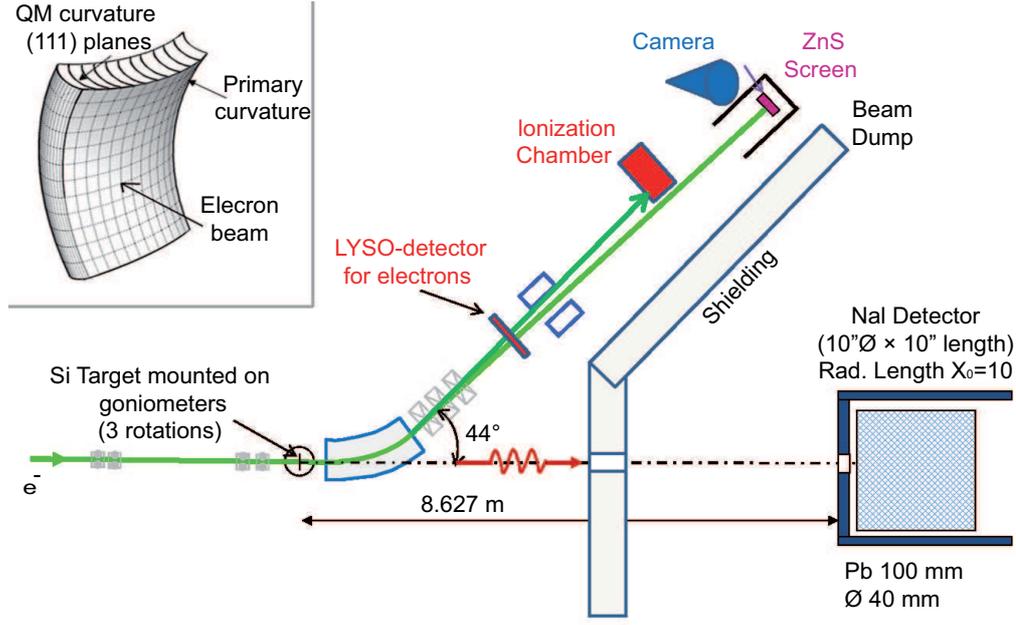}
    \caption{Scheme of the experimental setup at MAMI B. After the interaction with Si target mounted on a high precision goniometer, the 855 MeV electron beam is deflected horizontally by a bending magnet. A LYSO detector downstream of the crystal measures the beam distribution after interaction. Then a ionization chamber is positioned after a second dipole-magnet, which deflects particles along the vertical direction. It is employed to detect channeling. Just in front of the beam dump the beam spot can be monitored with a ZnS luminescent screen which is viewed by a CCD camera. The photon beam arrives at a NaI scintillator detector after 8.627 m from the crystal. The NaI detector is shielded by a 100 mm thick lead wall with a 40 mm opening to accept most of the photons. Top-left corner: sketch of the bent crystals. Primary curvature, quasi-mosaic curvature, and crystallographic orientation of the channeling planes are highlighted.} \label{Fig:setup}
\end{center}
\end{figure*}

The experimental setup at MAMI for both beam steering and radiation measurement is shown in Fig.\ref{Fig:setup}.
In particular, the MAMI B 855 MeV electron beam was used. Beam size and angular divergence in the plane of the crystal bending were 105 $\mu$m and 21 $\mu$rad, respectively. The divergence was considerably less than the Lindhard angle, which is 232 $\mu$rad for Si (111) and 274 $\mu$rad for Ge (111) for 855 MeV electrons. The samples were 15 $\mu$m thick, while their lateral size was optimized to suppress parasitic anticlastic deformation
across the central region of the crystal, resulting in a cylindrical surface \cite{GERMOGLI201581}. The beam entered the samples in the center of their largest surfaces, i.e. traversing 15 $\mu$m of material, as shown in the top-seft corner of Fig.\ref{Fig:setup}. The beam was aligned with the (111) planes, which were bent by the quasi-mosaic effect \cite{Ivanov.quasimosaicity,qmreview}.

Both crystals were mounted on a piezo-driven dynamical holder \cite{SYTOV2017,Salvador2018JINST}, which allowed to modify the sample curvature ($\theta_b$) without replacing the samples and without break the vacuum of the experimental apparatus. The holder was mounted on a high-precision goniometer with 5 degrees of freedom. The goniometer was used to align the crystals with the electron beam. The measurements were done at three different bending angles of the (111) planes for the Si sample (550, 750, and 1080 $\mu$rad) and three values of the (111) planes for the Ge sample (820, 1200, and 1430 $\mu$rad).

The electrons passed through the samples. Then, they were deflected by a dipole magnet onto a LYSO screen, which was used for the measurement of the deflection angle distribution. The direction of the deflection of the beam was orthogonal to the deflection imposed by the magnet. The crystal alignment was carried out using the signal of an ionization chambers installed downstream of the samples. A more precise alignment was reached via the analysis of the measured distribution of the deflection angle.

The measurements were carried out by considering different angular alignments between the samples and the beam ($\theta_x$):
\begin{itemize}
  \item Channeling alignment, $|\theta_x|<<\theta_L$;
  \item Volume reflection alignment, $-\theta_b<\theta_x<\theta_L$;
  \item Random alignment, $\theta_x>>\theta_L$;
  \item ``Anti volume reflection'' (AVR) alignment, i.e. opposite to the VR one, $\theta_L<\theta_x<\theta_b$.
\end{itemize}

\begin{figure*}
\begin{center}
    \includegraphics[width=1\textwidth]{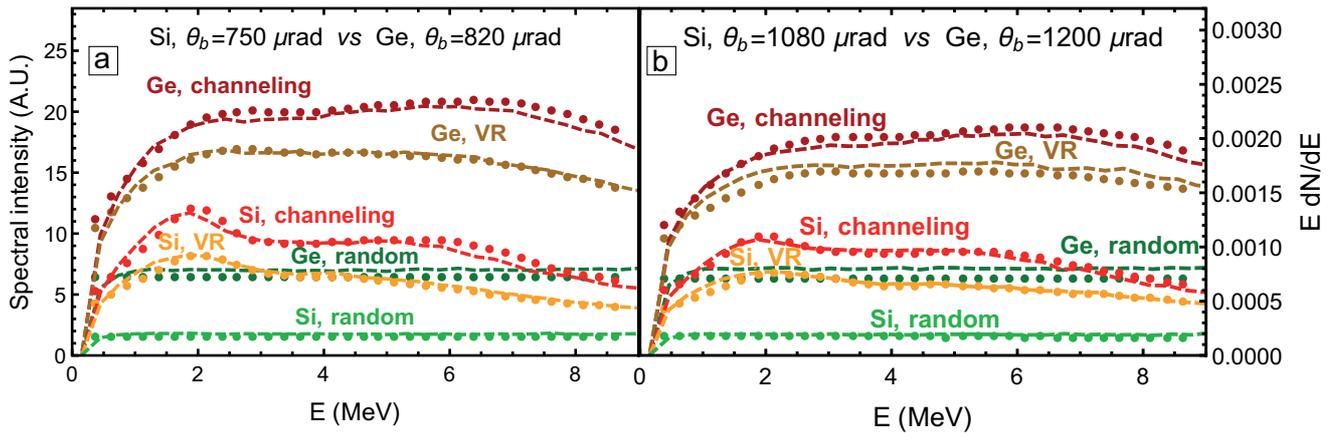}
    \caption{Experimental (lines) and simulated (points) radiation spectral intensities for channeling ($\theta_x \sim$0 $\mu$rad), volume reflection ($\theta_x = -\theta_b/2$) and random orientation ($\theta_x >$9 mrad) for both Si and Ge samples. (a) The bending angles were 750 $\mu$rad for Si and 820 $\mu$rad for Ge. (b) The bending angles were 1080 $\mu$rad for Si and 1200 $\mu$rad for Ge.} \label{Fig:radInt}
\end{center}
\end{figure*}

The emitted photons during the interaction of the beam particles with the samples were collected by a NaI detector, in front of which a collimator allowed to accept only photons emitted within the 40 mm aperture positioned at 8.627 m from the crystal (see Fig.\ref{Fig:setup}) \cite{Ban3,Lietti,MAMISetup}.

\section{Experimental results}

For the sake of comparison with the spectra usually displayed in the literature, the spectral intensities are obtained by multiplying the experimental spectra collected by the NaI detector by the photon energy. Fig. \ref{Fig:radInt} shows the measured radiation spectral intensities for Si and Ge samples, for comparable sample curvatures. The bending angles were 750 $\mu$rad for Si and 820 $\mu$rad for Ge in Fig.\ref{Fig:radInt}a and 1080 $\mu$rad for Si and 1200 $\mu$rad for Ge in Fig.\ref{Fig:radInt}b. Each of these measurements was carried out in channeling alignment ($\theta_x$ $\approx$ 0), in volume reflection alignment (at half of bending angle, i.e. $\theta_x$ $\approx$ $-\theta_b/2$), and at random alignment ($\theta_x$ $>$ 9 mrad).

\begin{figure*}
\begin{center}
    \includegraphics[width=0.95\textwidth]{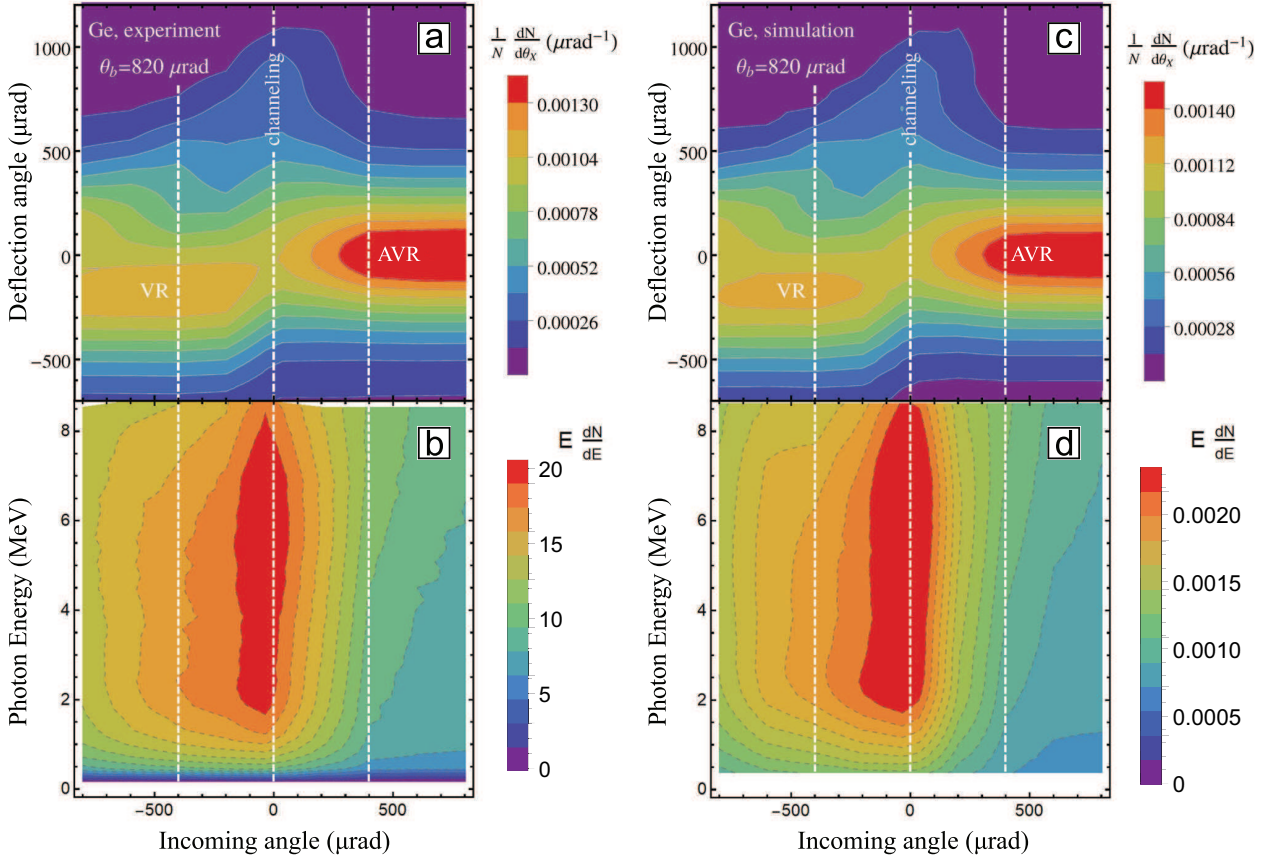}
    \caption{Experimental (left) and simulated (right) deflection angle distributions (top) and radiation spectral intensities (bottom) with respect to the alignment of the Ge bent crystal with bending angle of 820 $\mu$rad.} \label{Fig:def_rad}
\end{center}
\end{figure*}

As expected, the intensity of radiation generated in the Ge bent sample was higher than that for Si at similar crystal-to-beam alignment due to the higher atomic number $Z$ of Ge. For instance, radiation at random alignment was enhanced for the Ge sample w.r.t. the Si sample roughly by a factor of 4. Nevertheless, one may notice that the Si CR is more intense than that for random oriented Ge in the 0.5--8 MeV energy range thanks to the capability of channeling to enhance the radiation process. The Si CR is also more peaked than the Ge CR, which could be explained by a weaker spoiling contribution of multiple scattering to radiation \cite{Ban3} for a low-$Z$ material. On the other hand, both CR and VR radiation for the Ge sample were stronger than both CR and VR radiation for the Si one.
Since this is indeed the first measurement of radiation enhancement in a bent Ge crystal, we rotated the crystal from random to channeling orientation, passing through VR, in order to have a better insight on radiation characteristics.
For better comprehensibility, we have also plotted the distributions of the deflected beam as a function of the crystal alignment, which was already published in \cite{SYTOV2017}.
These plots demonstrate that, as already shown for Si in \cite{Ban3}, for a bent crystal the radiation intensity achievable in channeling condition is nearly maintained also for VR orientation (bottom plot of Fig.\ref{Fig:def_rad}) in the whole angular acceptance from $-\theta_b$ $<$ $\theta_x$ $<$ $\theta_L$.

\begin{figure}
\begin{center}
    \includegraphics[width=1\columnwidth]{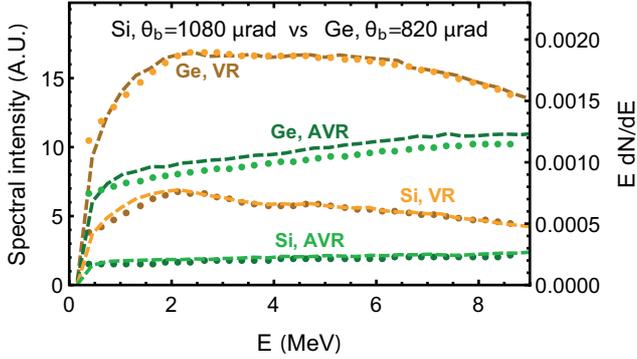}
    \caption{Experimental (lines) and simulated (points) radiation spectral intensities for volume reflection ($-\theta_b/2$) and anti volume reflection  orientation ($0<\theta_x<\theta_b$) for both Si and Ge bent crystals. Bending angles are 1080 $\mu$rad (Si) and 1430 $\mu$rad (Ge).} \label{Fig25}
\end{center}
\end{figure}

To highlight this peculiar behavior of bent crystals, we compared the radiation spectral intensities for VR orientation ($\theta_x \approx -\theta_b/2$) and for the opposite case of AVR, for which the crystal planes were misaligned w.r.t. the electron beam direction by an incoming angle of about $\theta_x \approx +\theta_b/2$. As an example, in Fig. \ref{Fig25} the direct comparison of the VR and AVR spectra for the bending angles of 1080 $\mu$rad and 820 $\mu$rad, for Si and Ge respectively, are presented. Here, one may notice a considerable enhancement of the radiation intensity at the VR alignment if compared to an identical incoming angle w.r.t. crystal planes in the opposite direction. Indeed, typically for straight crystals, the radiation intensity decreases quickly out of channeling condition, becoming harder while approaching to the coherent bremsstrahlung limit \cite{MAMISetup}.

\begin{figure*}
\begin{center}
    \includegraphics[width=1\textwidth]{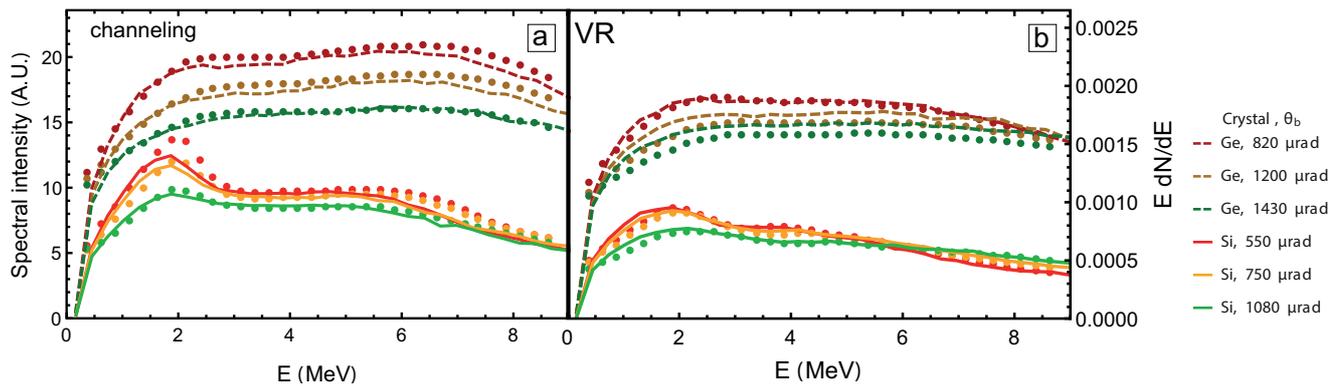}
\caption{Experimental (lines) and simulated (points) radiation spectral intensities for channeling (a) and volume reflection (b) alignments for both Si and Ge bent crystals with different values of bending angles. The VR spectra were measured for the incoming angle $\theta_x \approx - \theta_b/2$ except for the Si curvature $\theta_b$ = 550 $\mu$rad, for which both the measurement and the simulation were taken at $\theta_x \approx -0.7$ $\theta_b$.} \label{Fig3}
\end{center}
\end{figure*}

The similar intensity of VR radiation to CR allows one to generate rather intense radiation without very strict requirements on the crystal-to-beam alignment, while the beam divergence could be larger than the Lindhard angle, unlike in the case of channeling. Since the VR angular acceptance is equal to the bending angle of the crystal, in principle one may suppose to increase the angular acceptance for an intense gamma-source by increasing the crystal bending angle as much as needed. For this reason, it is crucial to determine the dependence of the VR radiation spectrum on the crystal bending angle. Fig. \ref{Fig3}a and \ref{Fig3}b display the channeling and VR spectral intensities for different crystal curvatures, respectively. CR was measured at three different values of bending angles $\theta_b$ for the Si crystal, namely 550, 750 and 1080 $\mu$rad and three values for the germanium one, i.e. 820, 1200 and 1430 $\mu$rad.
From Fig. \ref{Fig3}a it is evident that the main tendency for CR is the decrease in radiation intensity with an increase of the bending angle. 
This behaviour is due to the decrease of fraction of channeling particles for larger curvature, for which dechanneling contribution is stronger, i.e., the dechanneling length is shorter, as demonstrated in \cite{SYTOV2017}.
This behavior is confirmed for both Si and Ge crystals.



Fig. \ref{Fig3}b confirms the same tendency of the decrease of the radiation intensity with the increase of the bending angle also for VR radiation. Indeed, the more the crystal is bent, the greater the contribution of $\gamma$ emission due to interactions with the planes that are more misaligned with respect to the particle trajectory \cite{Ban2}. These photons are typically harder though radiation is less intense, resulting in a decrease of spectrum intensity.
Furthermore, due to the decrease of the critical angle for stronger bending angle, less particles are volume captured in channeling condition and, even if captured, are dechanneled faster. Indeed, it as been demonstrated in \cite{Ban3} that the volume captured particles are the main responsible for the maintenance of the CR peak also in VR orientation. Their decrease in number leads to a decrease of VR radiation peak.
Nevertheless, one may notice that the intensity of VR radiation remains comparable with the channeling case for all the values of bending radius considered for both Si and Ge bent crystals and the variation in its intensity depends slightly less on the curvature than in the case of CR.

Finally, one can conclude that the overall angular acceptance for the production of high intense radiation in a bent crystal is approximately equal to the crystal bending angle, $\theta_b$, being VR radiation of a similar intensity than CR and that the radiation intensity decreases with $\theta_b$ increase. This fact is quite evident in Fig.\ref{Fig:def_rad}, where it is shown that for the whole angular range from $\theta_x$ = 0 to $\theta_x$ = - $\theta_b$ = -820 $\mu$rad the intensity of radiation is maintained close to the CR peak.
One should pay by a small decrease of radiation intensity in order to provide considerably higher angular acceptance.

\begin{figure}
\includegraphics[width=1\columnwidth]{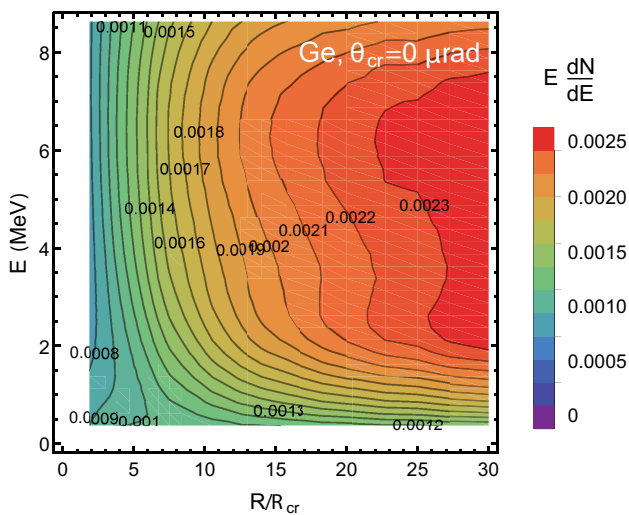}
\caption{Simulated radiation spectral intensities vs the ratio of the bending radius to its critical value for the Ge bent crystal.} \label{Fig4}
\end{figure}

\section{Monte Carlo simulation and discussion}


The interaction of the charged particles of the electron beam with the bent crystals and the generation of radiation were simulated with the CRYSTALRAD simulation code \cite{Sytov2019,Bandiera201544,SYTOV2015383}, which was already verified in several other experiments \cite{Wistisen2017,Ban2,Ban,Ban3,SYTOV2017,PhysRevLett.112.135503,Gui}. This code allows to simulate classical charged particle trajectories in straight and bent crystals, in the field of crystal planes or axes \cite{Lin,Biryukov} taking into account multiple and single Coulomb scattering on nuclei and electrons. The simulated spectral intensity E(dN/dE) vs. E, dN/dE being the photon emission probability, is calculated through the Baier-Katkov quasiclassical method\cite{BaierKatkov}. The CRYSTALRAD code takes into account the crystal geometry, such as anticlastic curvature, crystal torsion, miscut angle (namely the angle between the physical surface and the atomic planes of the crystal), and geometrical size of the crystal.


First of all, all the spectra presented in Fig.2--5 (dots) were well reproduced with the CRYSTALRAD simulation code.
Some discrepancy in the radiation intensity between simulation outcomes and experimental results may be inferred to one of the following reasons or their combination:
the chosen model of multiple and single Coulomb scattering, quantum corrections to Coulomb scattering at small angles \cite{PhysRevAccelBeams.22.054501}, and the influence of high-index planes in case of "random orientation".
Indeed, the weaker the scattering, the weaker the contribution of dechanneling of channeled and volume captured particles, i.e., the particles that mostly contribute to the radiation peak. At the same time, radiation in the case of random direction may be amplified by the presence of high-index skew planes, which may not have been perfectly avoided in choosing the random direction.

Nevertheless, the fairly good agreement between simulations and experiments permit us to exploit our Monte Carlo to investigate deeply the radiation generation in bent crystals. We focused our attention on the germanium case because its usage is a real novelty in the field and could open new possibilities for applications. In particular, we  investigated deeply the dependence of Ge Channeling Radiation spectra on curvature radius: Fig. \ref{Fig4} displays the radiation spectrum variation as a function of the ratio of the bending radius, R, w.r.t. its critical value $R_C$, namely the radius at which the potential well depth disappears and channeling is not possible for value smaller than $R_C$. This picture confirms that the CR intensity increases with the increase of the bending radius, i.e. decrease of the bending angle, $\theta_b$.
On the other hand, we have already shown in Fig.\ref{Fig:radInt}--\ref{Fig25} that also VR radiation decreases in intensity with $\theta_b$ increase, with a correspondent increase of the angular acceptance. In summary, one may find the best solution in terms of steering capability, radiation intensity, and angular acceptance for a particular application, such as intense radiation source or beam steerer, by tuning the bending angle of the crystal.

In general, crystals are very prospective radiation sources, since one can apply very intense electric fields naturally created.
The channeling intense radiation can be exploited in photoproduction experiments or to investigate nuclear structures, as already possible for CB radiation \cite{PhysRevC.93.065201}.

With a bent crystal, one can exploit the Volume Reflection radiation, which is not present for a straight crystal. This effect demonstrates lower but comparable radiation intensity with the channeling case, while the angular acceptance is considerably larger and tunable by proper choice of the crystal bending angle. Thus, VR in a bent crystal may resolve the problem of low angular acceptance of CR in a straight crystal, providing comparable intensity of radiation.

\section{Conclusions}

An experiment on channeling radiation and volume reflection radiation alignment has been carried out at MAMI B line, by studying the interaction of 855 MeV electrons with 15 $\mu$m thick silicon and germanium bent crystals.
The radiation spectral intensity of CR and VR has been measured and reproduced well using the CRYSTALRAD simulation code for three different values of the crystal curvature for the silicon crystal as well as three values for the germanium one. Both experiment and simulation outcomes show that the radiation intensity increased with the decrease of the bending angle, i.e. of the crystal curvature, for both CR and VR cases.

The intensities for germanium and silicon bent crystals with similar curvature were directly compared, demonstrating approximately a 2-factor increase in the radiation intensity for Ge in comparison with Si. Nevertheless, the Si CR is more peaked than the Ge CR, which could be explained by a weaker spoiling contribution of multiple scattering to radiation for a low-Z material. Finally, given its large angular acceptance, we fully demonstrated that VR radiation can be exploited for the realization of intense radiation sources with poor emittance beam also using higher Z material than Si or C.

The results presented in this paper are relevant for the generation of X-ray and $\gamma$ radiation in bent crystals, including the crystals of higher-Z materials and crystalline undulators \cite{korol2013channeling}. Moreover, Monte Carlo simulations provide the possibility of prediction of the results for future experiments, which is also essential for the development of the innovative radiation sources.

\begin{acknowledgements}

We acknowledge partial support of the INFN through the CSN5-ELIOT experiment, the MC-INFN and the Grant73-OSCaR projects. We also acknowledge partial support by the European Commission through the N-LIGHT Project, GA 872196. M. Romagnoni acknowledges partial support from the ERC Consolidator Grant SELDOM G.A. 771642. We also acknowledge the CINECA award under the ISCRA initiative for the availability of high performance computing resources and support. We acknowledge Professor H. Backe for fruitful discussion and M. Rampazzo, A. Pitacco, and A. Minarello for technical assistance in dynamic holder realization.

\end{acknowledgements}

\bibliographystyle{unsrt}
\bibliography{RADMAMI2016_arxiv}

\end{document}